\documentclass{article}

\usepackage{arxiv}

\usepackage[utf8]{inputenc} 
\usepackage[T1]{fontenc}    
\usepackage{hyperref}       
\usepackage{url}            
\usepackage{booktabs}       
\usepackage{amsfonts}       
\usepackage{nicefrac}       
\usepackage{microtype}      
\usepackage{graphicx}
\usepackage[numbers]{natbib}
\usepackage{doi}

\usepackage{siunitx}
\usepackage{caption}
\usepackage{subcaption}
\usepackage{amsmath,amsfonts,amssymb}
\usepackage{graphics}
\usepackage{amssymb}
\usepackage{bm}

\title{Memory effects in friction: the role of sliding heterogeneities}

\author{ Vincenzo Fazio$^{1}$, Vito Acito$^{1}$, Fabien Amiot$^{2}$, Christian Fr\'etign$^{1}$y and Antoine Chateauminois$^{1,*}$ \\	
	$^{1}$~\textit{Soft Matter Science and Engineering Laboratory (SIMM)} \\
	PSL Research University, UPMC Univ. Paris 6, Sorbonne Universit\'es, ESPCI Paris\\
	 CNRS, 10 rue Vauquelin, 75231 Paris Cedex 05, France\\		   
		$^{2}$~\textit{Univ. Bourgogne Franche-Comt\'e, Institut FEMTO-ST, CNRS/UFC/ENSMM/UTBM} \\
		 \textit{Département M\'ecanique Appliqu\'ee} \\
		 24 rue de l'Epitaphe, 25000 Besançon, France\\		 
	\texttt{* antoine.chateauminois@espci.fr} \\
}

\renewcommand{\undertitle}{ }

\begin{document}
\maketitle

\begin{abstract}
We report on memory effects involved in the unsteady state frictional response of a contact interface between a silicone rubber and a spherical glass probe when it is perturbed by changes in the orientation of the driving motion or by velocity steps. From measurements of the displacement fields at the interface, we show that observed memory effects can be accounted for by the non-uniform distribution of the sliding velocity within the contact interface. As a consequence of these memory effects, the friction force may no longer be aligned with respect to the sliding trajectory. In addition, stick-slip motions with a purely geometrical origin are also evidenced. These observations are adequately accounted for by a friction model which takes into account heterogeneous displacements within the contact area. When a velocity dependence of the frictional stress is incorporated in this the model, unsteady state regimes induced by velocity steps are also adequately described. The good agreement between the model and experiments outlines the role of space heterogeneities in memory effects involved in soft matter friction.\\
\end{abstract}

\keywords{transient friction \and rubber \and velocity jumps \and curvilinear motions}
\section{Introduction}
Transient frictional regimes pertain to many practical situations encountered in everyday life. As typical examples, one can mention the transition of a contact from rest to steady-state sliding or the wide variety of contact instabilities, including stick-slip motions, which can be encountered up to the earthquake scale. Early experiments carried out on rocks by geophysicists~\cite{dietrich1972,dietrich1978,dietrich1979} or on metals by Rabinowitcz \cite{rabinowicz} demonstrated that the time-dependent changes in the friction force during transient regimes involve complex history effects. A typical situation explored by these studies is the frictional response of a contact when the slip rate is changed suddenly from one value to another greater value: a positive jump in the frictional stress followed by a long-term decay to steady-state over a characteristic length-scale is then observed. In order to describe these observations, Rice and Ruina~\cite{rice1983,ruina1983} have developed a seminal constitutive law where the friction force is dependent on slip rate and on phenomenological state variables accounting for the fading memory of the contact. In this model, the state variables basically reflect the internal degrees of freedom of the sliding system.

At the microscopic scale, the underlying physical mechanisms behind the state-and-rate friction laws have mostly been addressed within the context of rough, multi-contact, interfaces~\cite{baumberger_solid_2006,Bureau2002,heslot1994,berthoud1999,bendavid2010,putelat2011}. In these approaches, the history dependence of friction is ascribed to contact area ageing and rejuvenation as a result of creep and slippage mechanisms at the scale of micro-asperity contacts.

Noticeably, state-and-rate models mostly consider extended contact interfaces, where the non uniformity of the deformation of the contacting bodies is generally discarded. In finite size contacts, deformation gradients are however invariably induced at the contact scale. Such gradients are expected to be especially strong in the case of finite size contacts between soft substrates such as rubbers or gels where frictional shear stresses are typically of the order of magnitude of the shear modulus~\cite{chateauminois2017}. As a consequence, the non uniform deformation field of the contacting bodies should be specified to describe the state of the system. Any approach ignoring these degrees of freedom - as it is the case in most macroscopic friction models -may thus fail in the description of history effects involved in transient regimes.

In this study, we tackle memory effects in friction from the perspective of the transient sliding heterogeneities which result from the deformation of a finite size contact area during unsteady state sliding regimes. For that purpose, a smooth, single-asperity, contact interface between a deformable rubber substrate and a rigid spherical probe is perturbed by the application of either non rectilinear sliding motions or a velocity step. In the case of non rectilinear motions, we show from measurements of the displacement fields within the contact that stress and strain heterogeneities keep a memory of the past trajectories. At the macroscopic scale, one of the consequences of this memory effect is the development of friction force components normal to the sliding trajectories. In such unsteady state situations, we show that the observed transient regimes correspond to the \textit{characteristic sliding distance} which is needed to the contact to recover from a trajectory or sliding velocity perturbation.\\ 

In the case of rigid surfaces with anisotropic frictional properties and/or curved sliding paths, the existence of directional effects in friction have previously been reported both theoretically~\cite{zmitrowicz1999a,Zmitrowicz1999b,zmitrowicz2006} and experimentally~\cite{halaunbrenner1960,rabinowitcz1957,chateau2013,tapia2016}. We show here that they can also be a consequence of the loss of symmetry resulting from the transient contact deformations induced by changes in the orientation of the sliding motion.  We also show that another feature of non rectilinear sliding paths is the possible occurrence of new kind of stick-slip motions with a purely geometric origin. Differently from classical stick-slip motions induced by the coupling between the constitutive friction law and the dynamics of the system~\cite{persson2000book,szlufarska2008}, discontinuous sliding motions are here induced by the curvature of the trajectory.\\

As a starting point, we first consider the simplified, theoretical, situation of a point contact driven by an isotropic spring at a constant velocity along a linear trajectory which undergoes a sudden change in direction. Broken line sliding experiments with finite size contacts between a silicone elastomer and a spherical glass probe are subsequently discussed in section 3 in the light of this point contact toy model. From the analysis of the orientation of the macroscopic friction force and of the sliding velocity within the contact, we formulate in section 4 a friction model based on the assumption that the interfacial shear stress is oriented  along the local interfacial velocity. In section 5, this model is extended to the description of experimental results for various trajectories (broken lines, circles, sine waves). In a last section, this discussion is declined in the case of a velocity step. Full details regarding the friction devices and experimental conditions are provided in a Method section at the end of the manuscript.\\  
\section{A primer: the point contact as a toy model}
As a toy model, we first consider a point contact which is held in the laboratory frame by means of a system including a 2D isotropic stiffness $k$. The point contact is lying on a flat substrate which is driven at a constant velocity along a linear trajectory which undergoes a sudden change of direction. We assume that inertial forces can be neglected and that friction obeys a standard model: the magnitude of the friction force applied to the sphere $\mathbf{F}_T$ remains constant in the sliding regime : $\left|\mathbf{F}_T(t)\right|=T$. Lower values of the friction force correspond to stick phases, where the relative velocity between the slider and the substrate vanishes. 

Denoting respectively $\mathbf{R}(t)$ and $\mathbf{r}(t)$ the positions of the holder and of the contact point in a frame attached to the moving substrate, the equilibrium condition for the sphere reads
\begin{equation}
\mathbf{F}_T(t)=-k\left( \mathbf{r}(t)-\mathbf{R} (t)\right).
\end{equation} 

\textit{In the sliding regime}, as the magnitude of the friction force is constant, we have $\left|\mathbf{r}(t)-\mathbf{R}(t)\right|=\lambda$, where we introduce  a \textit{tribo-elastic} length $\lambda=T/k$. Then, in this regime, the distance between the slider and the driving point is $\lambda$. It turns out that the slider follows a tractrix curve on the moving substrate, \textit{i.e.} a curve with the property that the distance from any point on the curve to a given line, measured along the tangent of the curve, is constant (see Supplementary Information for more details). This is, for example, the curve followed by the  back wheel of a bike for a prescribed trajectory of the front wheel. Tractrix have a very rich mathematical history. It appears in many textbook on differential geometry of curves and a concise history can be found in the introduction of reference~\cite{cady1965}, for example.  

\textit{In the stick regime}, the distance between the slider and the driving point is less than the characteristic length $\lambda$ and the position of the slider remains fixed on the substrate plane, $\mathbf{r}(t)=\mathbf{r}_S$. This stick phase lasts until the slider-driving point distance reaches $\lambda$ again. Passed this point, the slider follows a tractrix with $\mathbf{r}_S$ as initial condition.

To determine the condition for sliding, one may express that, for thermodynamics reasons, the work per unit of time paid by the driver against sliding friction should be positive $\mathbf{F}_T(t)\cdot\frac{d\mathbf{R}}{dt}>0$, or
\begin{equation} 
\frac{d\mathbf{R}}{dt}\cdot\left(\mathbf{R}(t)-\mathbf{r}(t)\right)>0.
\end{equation} 
This sliding condition expresses that the angle between the velocity of the moving substrate and the  vector joining the slider to the holder should be less than $\pi/2$. 

\begin{figure}
\centering
\begin{subfigure}[b]{1\textwidth}
	\centering
	\includegraphics[width=0.4\textwidth]{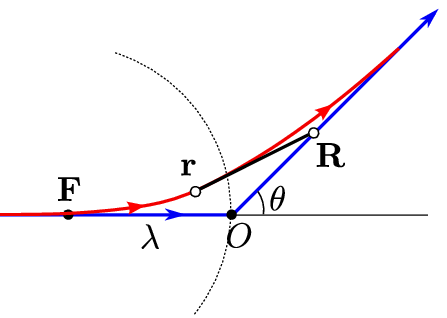}
	\caption{}
	\label{fig:angles}
\end{subfigure}
\hfill
\begin{subfigure}[b]{1\textwidth}
	\centering
	\includegraphics[width=0.5\textwidth]{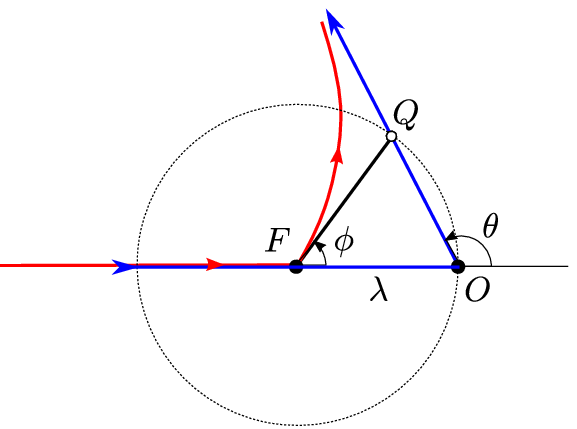}
	\caption{}
	\label{fig:phig}
\end{subfigure}
\caption{Trajectories of a point contact when the moving substrate follows the horizontal axis until point $O$ where it suddenly changes its direction (blue line). \textit{(a)}  When the reorientation angle $\theta<\pi/2$, the slider (at position $\mathbf{r}$) remains at a distance $\lambda$ from the holder (at position $\mathbf{R}$) and continuously slides along a smooth path which is a portion of a tractrix passing through point $F$ (in red).  \textit{(b)} When $\theta>\pi/2$, the slider stops at the point $F$ at a distance $\lambda$ from the reorientation point $O$ and stays there until the slider reaches the point $Q$. Its trajectory is then a portion of a tractrix which passes through $F$.}
\label{fig:point_contact}
\end{figure}

When the substrate follows a broken line, then two situations may occur. When the reorientation angle is less than $\pi/2$, the slider remains in a sliding regime and it follows a portion of a tractrix curve (figure~\ref{fig:angles}). In the opposite case, the slider stops, as its distance from the holder becomes less than $\lambda$ and it slides again when this spacing reaches $\lambda$ again (figure~\ref{fig:phig}). Then, it follows a portion of a tractrix (see Supplementary Information for more details).

\begin{figure}[!h]
\centering\includegraphics[width=0.8\textwidth]{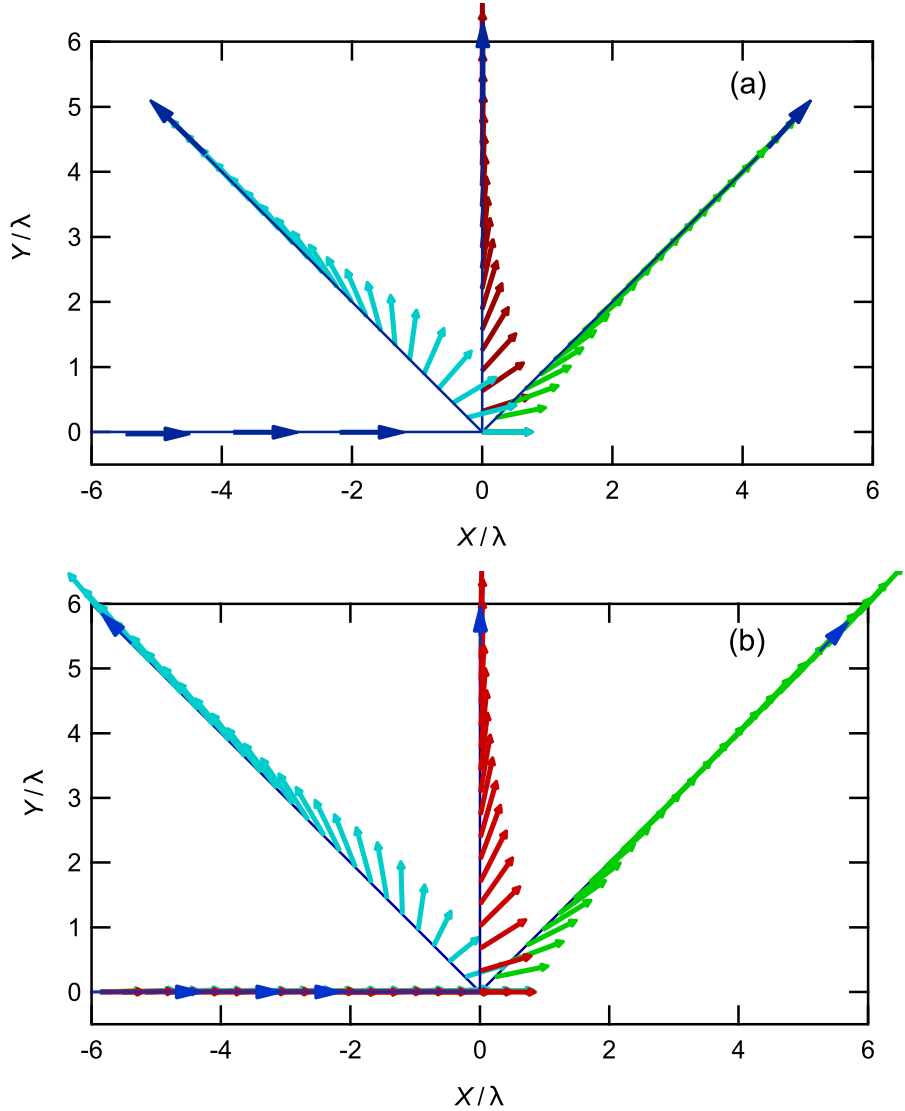}
\caption{Vector plot of the friction force along a piecewise rectilinear motion where the orientation of the imposed linear motion is changed by an angle $\theta$. (a) Calculated friction force applied to a point contact by the substrate; (b) Experimental friction force applied to the glass lens by the PDMS substrate ($V=0.1$~\si{\milli\metre\per\second}). Green: $\theta=\pi/4$; red: $\theta=\pi/2$; light blue: $\theta=3\pi/4$. From the measurement of the steady-state friction force under linear sliding conditions, $F_S=1.39 \pm 0.03\:\si{\newton}$  and of the lateral contact stiffness, $k_c=4.73 \pm 0.07 \:10^{3}\: \si{\newton\per\meter}$ the value of the elastic length $\lambda=F_s/k_c$ is found to be $0.29 \: \si{\milli\meter}$. The blue lines and arrows correspond to the imposed ($X,Y$) trajectories.}
\label{fig:fig_L_force_vector}
\end{figure}

In figure~\ref{fig:fig_L_force_vector}a, a vector plot of the calculated friction force applied to the point contact is presented, before and after a sudden change of the direction of the imposed linear motion of the substrate for three different reorientation angles. A transient domain is observed where the friction force gradually changes from the initial orientation to the new one. In this region, the friction force is not tangent to the imposed trajectory.

More generally, for any driving curve, the slider trajectory is then analogous to that of a dog pulled by its master at the end of a leash. The dog stops as soon as the leash is loosened. The qualitative characteristics of the dog trajectories highly depend on the relative values of the leash length and of the geometrical characteristics of the driving curve. For highly curved  driving trajectories, stick and slip phases may also alternate.

In real friction situations with a finite contact size, even with a rigid driving device, an isotropic stiffness exists in the system which is related to the elasticity of the contacting materials. One may therefore expect the same qualitative behaviours as in the point contact case to be observed. However, the details should differ since the displacements and stress fields may not be uniform in the contact area. In the following, experiments are presented for a glass lens in contact with a rubber substrate driven along different paths.\\

%
%
\section{Broken line linear sliding experiments}
\begin{figure}[!ht]
\centering
\includegraphics[width=0.7\textwidth]{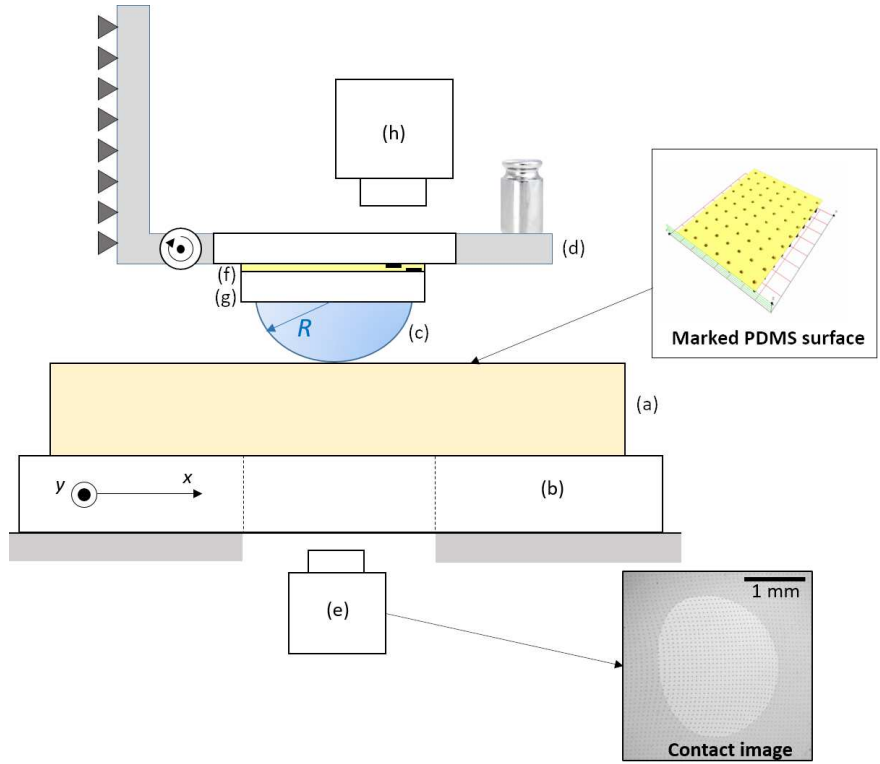}
\caption{Schematic description of the custom-built setup for friction measurements under non rectilinear sliding motions. A surface-marked PDMS substrate (a) is fixed on two crossed, motorized, linear translation stages (b) allowing to vary the sliding direction. Contact with a plano-convex glass lens of radius $R=$12.96~$\si{\milli\meter}$ (c) is achieved under imposed normal load using a dead weight arm (d).  A CMOS camera (e) allows contact visualization through the transparent PDMS substrate. The frictional force components in the contact plane are measured using a custom made sensor consisting in a silicone disk (f) enclosed between a glass disk (g) and a glass plate patterned on their internal surfaces (patterned areas are indicated by the thick marks). After stiffness calibration, the force is determined from sub-pixel measurements of the relative displacement between the two patterns using a CMOS camera (h).}
\label{fig:setup}
\end{figure}

We now consider finite size contacts between a smooth silicone substrate and a smooth glass lens under imposed normal load conditions. Here, an elastic length $\lambda$ arises from the deformation of the PDMS substrate. It may be defined as $\lambda=F_S/k_c$, where $F_S=\left|\mathbf{F}_S\right|$ is the magnitude of the steady-state friction force under linear sliding and $k_c$ is the lateral contact stiffness. Here, $k_c$ was determined experimentally from the initial linear part of the force versus displacement curves, \textit{i.e.} when the static contact of radius $a_0$ is dragged on the surface of the PDMS substrate in the absence of any significant slip at the interface~\cite{Mindlin1953}.\\
Using the friction set-up detailed in \ref{fig:setup} (for further details, see the Methods section), a sudden change $\theta$ in the orientation of the linear sliding trajectory of the PDMS substrate is applied after the contact has been prepared in a steady-state frictional state at a constant velocity $V$. The latter is achieved by moving the PDMS substrate over a distance of 2~$\si{\milli\meter}$, i.e. larger than the initial static contact radius ($a_0=1.38 \pm 0.01 \si{\milli\meter}$).
Figure~\ref{fig:fig_L_force_vector}b displays vector plots of the friction force $\mathbf{F}_T$ applied to the glass lens for various positions ($X(t),Y(t))^\intercal$ of the moving PDMS substrate driven at a velocity $V=0.1 \si{\milli\metre\per\second}$. In this figure, the blue lines correspond to the trajectories of the PDMS substrate for $\theta=\pi/4,\:\pi/2$ and $3\pi/4$, the change $\theta$ in their orientations being applied at the position ($X=0,Y=0$). Immediately after the change in the orientation of the driving motion, the friction force is no longer collinear to the sliding trajectory. Then, a progressive realignment of the friction force with respect to the sliding direction is observed, in a way which is qualitatively similar to the calculated point contact situation (cf figure~\ref{fig:fig_L_force_vector}a).

\begin{figure}[!h]
\centering\includegraphics[width=11cm]{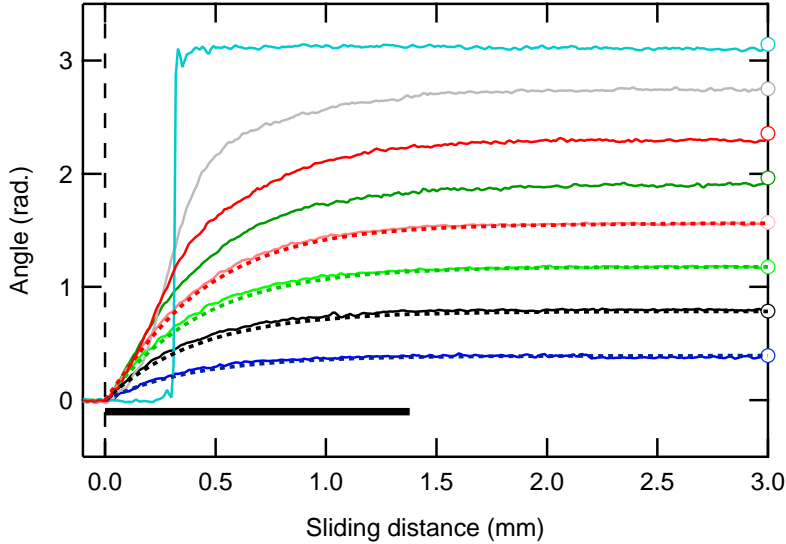}
\caption{Orientation of the friction force after a change in the direction of the imposed linear motion by an angle $\theta$ with respect to the initial sliding motion ($V=0.1$~\si{\milli\metre\per\second}). From bottom to top: $\theta=\pi/8,\:\pi/4,\:3\pi/8,\:\pi/2,\:5\pi/8,\:3\pi/4,\:7\pi/8$ and $\pi$. Open symbols correspond to the prescribed change $\theta$ in the direction of the sliding motion. The radius of the static contact is indicated by the length of the horizontal bold line. Dotted lines corresponds to simulations carried out using equation~(\ref{eq:ligne_brisee}) in situations without stick ($\theta\leq \pi/2$).}
\label{fig:fig_L_force_angle}
\end{figure}

In figure~\ref{fig:fig_L_force_angle}, the angle of the friction force with respect to the $X$ axis of the sliding motion is reported as a function of the sliding distance for values of the angle $\theta$ ranging from $\pi/8$ to $\pi$. Except for $\theta=\pi$, it turns out that the reorientation of the friction force occurs over a length close to the size of the static contact radius (shown as a horizontal bold line in the figure). Additional experiments carried out at $V=10 \: \si{\micro\meter\per\second}$ (not shown) indicated that this length is not significantly affected by the sliding velocity. For $\theta=\pi$, the motion is fully reversed with the friction force passing through zero; as a consequence, the direction of the friction force switches instantaneously from 0 to $\pi$.

\begin{figure}[!h]
\centering
\begin{subfigure}[b]{1\textwidth}
	\centering
	\includegraphics[width=10cm]{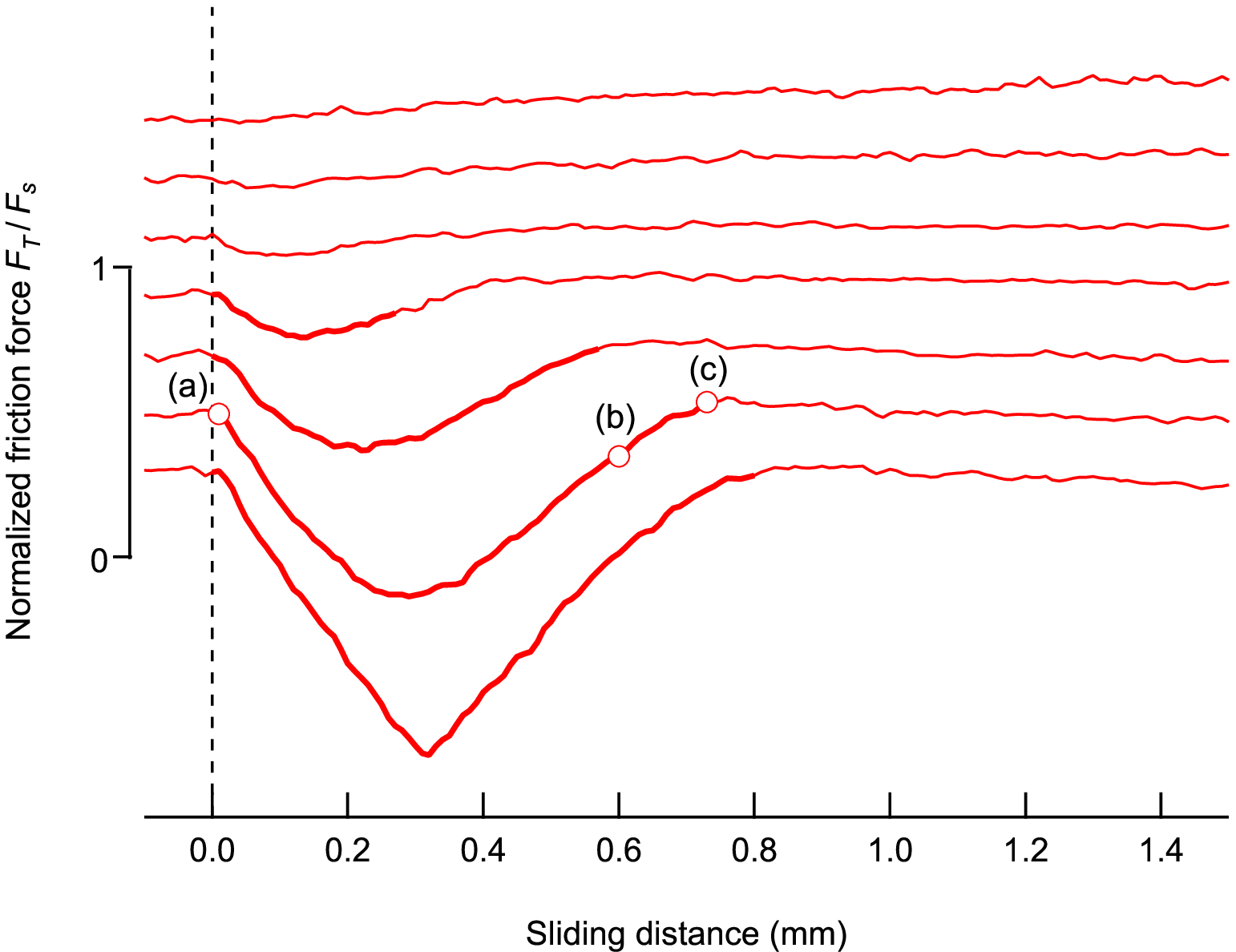}
	\label{fig:fig_L_force_magnitude}
\end{subfigure}
\hfill
\begin{subfigure}[b]{1\textwidth}
	\centering
	\includegraphics[width=12cm]{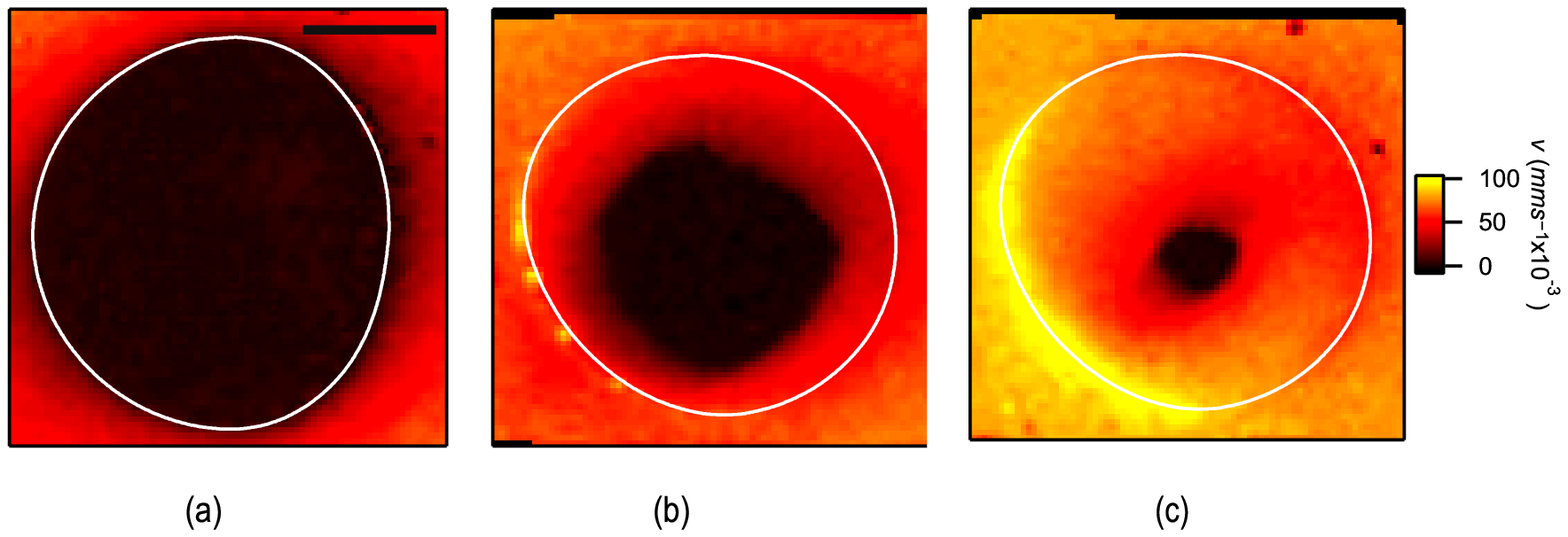}
	\label{fig:fig_L_stick}
\end{subfigure}
\caption{Top: Traces of the normalized friction force $F_T/F_S$, where $F_S$ is the steady-state friction force, after a change at in the direction of the imposed linear motion by an angle $\theta$ with respect to the initial sliding motion ($V=0.1$~\si{\milli\metre\per\second}). The dotted line indicates the position at which the change in the angle of the sliding direction is applied. From top to bottom: $\theta=\pi/4,\:3\pi/8,\:\pi/2,\:5\pi/8,\:3\pi/4,\:7\pi/8$ and $\pi$. Bold lines indicate the existence of stick area within the contact. Bottom: magnitude of the sliding velocity within the contact area (delimited by the white line) at three different stages (indicated by (a), (b) and (c) in the top graph) during a stick event for $\theta=7\pi/8$. The black horizontal bar correspond to 1~$\si{\milli\meter}$.}
\label{fig:fig_L_force}
\end{figure}

Figure~\ref{fig:fig_L_force} (top) shows the magnitude of the friction force $\mathbf{F}_T$ during the transient regime for $\pi/4 \leq \theta \leq \pi$. In order to account for slight variations in the magnitude of $\mathbf{F}_T$ between different experiments, the friction force has been normalized with respect to its steady-state value $F_S$. When $\theta$ is increased above $\pi/2$, a transient decrease in the magnitude of the friction force $F_T$ is evidenced whose amplitude increases with $\theta$. From the examination of the sliding velocity fields within the contact (figure~\ref{fig:fig_L_force}, bottom), it turns out that this drop in $\mathbf{F}_T$ is associated to a re-stick of the contact immediately after the orientation of the sliding trajectory has been changed. Indeed, the slope of the force versus displacement curve immediately after the change in the orientation of the trajectory would correspond to the lateral contact stiffness $k_c$ of the sheared contact. As the PDMS substrate is further displaced with respect to the lens, slip progressively re-invades the contact from its periphery, until a full sliding condition is achieved close to the point where the magnitude of the friction force recovers its steady state-value (in figure~\ref{fig:fig_L_force} (top), the occurrence of such a partial slip condition is indicated by bold lines). The development of slip within the contact results in a continuous decrease in the contact stiffness which would make a quantitative comparison with the point contact model difficult.\\
However, similarly to the point contact situation, the finite size contact re-stick for a critical value of the angle $\theta$ close to $\pi/2$. This analogy between point and finite size contacts regarding the slip-to-stick condition can be accounted for by the fact that, at the time the sudden change $\theta$ in the orientation of the trajectory is applied, the velocity field is very uniform, except for small Poisson's effects~\cite{Nguyen2011}. The contact is thus characterized by a nearly unique orientation of the sliding velocity, irrespective of its size: this is equivalent to a point contact at the position where it undergoes a change in the sliding direction. 

At this stage, it turns out that the simple point contact model and the associated tractrix can provide a qualitative understanding of the change in the orientation and of the magnitude of the friction force following a discontinuity in the sliding direction. However, we show below that the details of the orientation of the friction force during the transient regime depend on the heterogeneous sliding conditions which are achieved at the contact interface as a result of the deformation of the soft substrate.

\begin{figure}[!h]
\centering\includegraphics[width=1\textwidth]{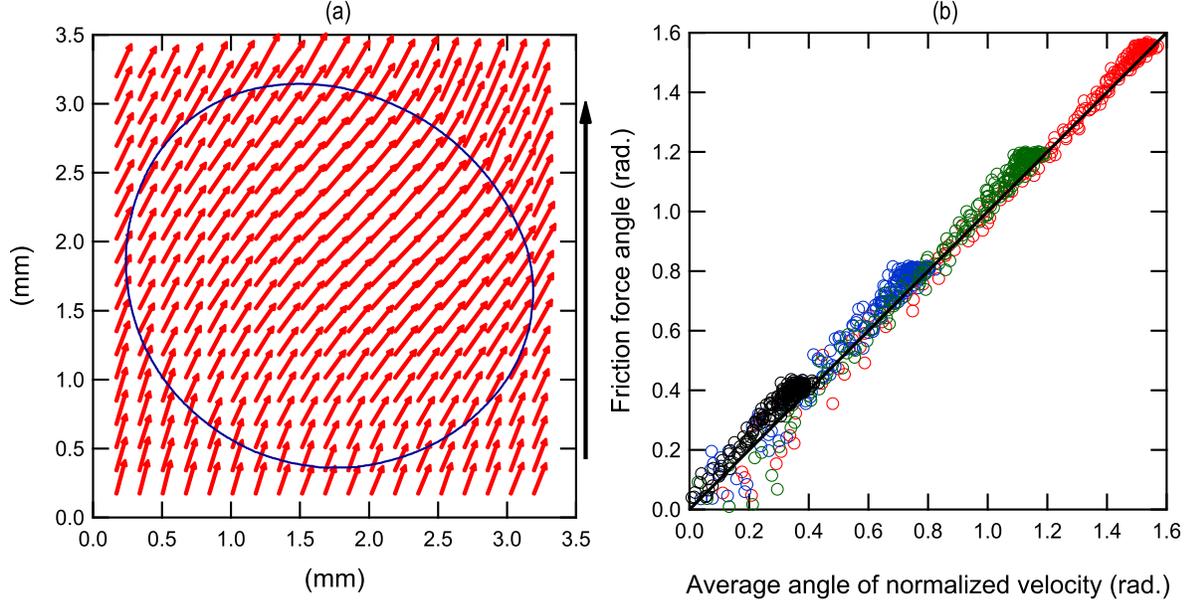}
\caption{Orientation of the velocity field at the surface of the PDMS substrate after a change  $\theta$ in the direction of the imposed motion ($V=0.1$~\si{\milli\metre\per\second}). (\textit{a}) Vector plot of the surface velocity field for $\theta=\pi/2$ and a sliding distance of 0.5~$\si{\milli\meter}$ following the change $\theta$. The blue line delimits the contact area. The PDMS substrate is moved vertically from bottom to top as indicated by the arrow. (\textit{b}) Orientation of the macroscopic friction force $\boldmath{F}_T$ as a function of the average orientation of the normalized sliding velocity $\mathbf{V}/\left|\mathbf{V}\right|$ within the contact. Black: $\theta=\pi/8$; blue: $\theta=\pi/4$; green: $\theta=3\pi/8$; red: $\theta=\pi/2$. The black line correspond to a unit slope.}
\label{fig:fig_L_field}
\end{figure}

For that purpose, we focus on the relationship between the orientation of the macroscopic friction force and the distribution of the sliding velocity within the contact for $\theta \leq \pi/2$, i.e. in the absence of stick. As shown in figure~\ref{fig:fig_L_field}a, measurements of the displacements at the surface of the substrate reveal an heterogeneous distribution of the velocity field. It also appears that, on an average, the local sliding velocity is strongly misaligned with respect to the imposed motion.

In figure~\ref{fig:fig_L_field}b, the orientation of the macroscopic friction force $\mathbf{F}_T$ is reported as a function of the average orientation of the normalized sliding velocity $\mathbf{V}/\left|\mathbf{V}\right|$ for $\theta \leq \pi/2$. It turns out that the orientation of the macroscopic friction force matches perfectly the average orientation of the sliding velocity field within the contact. This observation can be justified by considering that (\textit{i}) locally, the frictional shear stress $\mathbf{\tau}$ is tangent to the sliding direction; (\textit{ii}) $\mathbf{\tau}$ is independent on contact pressure. This latter condition is supported by previous results using similar smooth single asperity glass/PDMS~\cite{Nguyen2011,chateauminois_local_2008}. In what follows, we derive a model based on these two assumptions in order to account for the progressive re-orientation of the velocity field and for the macroscopic friction force during the transient regime.\\
%
%
\section{Unsteady-state friction model}

In this section, a model is derived to account for the experimental results, by assuming only that, within the contact area, the interfacial shear stress is oriented along the \textit{local interfacial velocity}, the velocity field being itself shaped by the friction history. In accordance with earlier experimental results, the amplitude of the friction stress for this smooth glass-rubber system is supposed to be independent of the normal stress and (weakly) dependent of the interfacial velocity~\cite{Nguyen2011,chateauminois_local_2008}. The model is implemented within the framework of the linear elasticity approximation, neglecting inertia effects. To simplify, the Poisson's ratio of the substrate is taken as $\nu=1/2$ as it is nearly the case for rubber materials. While experimental observations show that the contact is not perfectly circular during sliding (cf figure~\ref{fig:fig_L_field}a), it is assumed for the sake of simplicity that the contact is a disk.

A spherical lens, fixed in the laboratory frame, is maintained in contact with a rubber substrate which is displaced in its plane. Its trajectory $\mathbf{R}(t)$ is prescribed and the problem is to describe the displacements and the friction stress $\mathbf{u}=(u_x,u_y)^\intercal$ and $\mathbf{\boldsymbol{\tau}}=(\tau_{xz},\tau_{yz})^\intercal $ in the contact area to deduce the global friction force $\mathbf{F}_T$ which applies to the lens. The velocity dependence of the frictional stress reads
\begin{equation}
\boldsymbol{\tau}=-\tau_0(\left|\mathbf v\right|)\frac{\mathbf{v}}{\left|\mathbf v\right|},
\label{eq:frictlaw}
\end{equation}
where $\mathbf{v}$ is the interfacial velocity and $\tau_0(\left|\mathbf v\right|)$ is a prescribed friction law which, as a first approximation, can be deduced as the ratio of the friction force to the contact area in steady state rectilinear friction experiments. In order to evaluate the interfacial velocity, one may remark that a point $\boldsymbol{\rho}$ which is fixed in the lens frame is, at a time $t$, in contact with a point of the substrate, which would be at $\boldsymbol{R}_0(t)$ at rest in the substrate frame and which underwent a displacement $\mathbf{u}(t,\mathbf{R}_0(t))$ under the influence of the stress field: $\mathbf{R}_0(t)+\mathbf{u}(t,\mathbf{R}_0(t))=\mathbf{R}(t)+\boldsymbol{\rho}$. The displacement velocity is, up to the first order,
\begin{equation}
\frac{d\mathbf u}{dt}=\frac{\partial \mathbf u}{\partial t}+\left( \frac{\partial \mathbf R}{\partial t}\cdot\boldsymbol\nabla  \right)\mathbf u.
\end{equation}
The relative velocity of this point with respect to the lens is then
\begin{equation}
\mathbf v=-\mathbf V+\left[\frac{\partial }{\partial t}+\left(\frac{\partial \mathbf R}{\partial t}\cdot\boldsymbol\nabla  \right)\right]\mathbf u.
\label{eq:relative_vel}
\end{equation}
In the case of incompressible materials such as rubber, Green's analysis establishes that the static surface displacements induced by the vertical and lateral components of a point loading on an elastic half-space are fully decoupled~\cite{landau1986}:
\begin{equation}
u_i=G_{ij}*\tau_{j,z},
\end{equation}
where the symbol $*$ stands for a 2D convolution operation and
\begin{align}
G_{xx}&=\frac{3}{4 \pi E}\left(\frac{1}{r}+\frac{x^2}{r^3}\right),\\
G_{yy}&=\frac{3}{4 \pi E}\left(\frac{1}{r}+\frac{y^2}{r^3}\right),\\
G_{xy}&=G_{yx}=\frac{3}{4 \pi E}\left(\frac{xy}{r^3}\right).
\end{align}
With a matricial notation, one may express
\begin{equation}
\mathbf{u}=G*\boldsymbol \tau
\end{equation}
and thus
\begin{equation}
\mathbf v=-\mathbf V+\left[\frac{\partial }{\partial t}+\frac{\partial \mathbf R}{\partial t}\cdot\boldsymbol\nabla  \right]G*\boldsymbol \tau.
\label{eq:interfvel}
\end{equation}
The local friction hypothesis equation~(\ref{eq:frictlaw}) together with the above expression give a self-consistent problem, which contains a time differential equation for the traction field $\boldsymbol\tau$. The non-linearity of the friction law equation~(\ref{eq:frictlaw}) makes it difficult solve the problem in a closed form. In the following, the equation is numerically solved by evaluating the above equation in the Fourier space:
\begin{equation}
\mathbf {\hat v}=-\mathbf {\hat{V}}+\left[\frac{\partial }{\partial t}+2i\pi\left( V_xk_x+V_yk_y \right)\right]\hat G\hat{\boldsymbol \tau },
\end{equation}
where $\hat{f}(\mathbf{k})=\iint f(\mathbf{r})\exp\left( 2i\pi\mathbf{k}.\mathbf{r} \right)\,d^2\mathbf{r}$, $(V_x,V_y)^\intercal=\left( dR_x/dt,dR_y/dt\right)^\intercal$ and
\begin{equation}
\hat{G}=\frac{3}{4\pi E}\begin{pmatrix} \frac{1}{k}+\frac{k_y^2}{k^3}&-\frac{k_xk_y}{k^3} \\-\frac{k_xk_y}{k^3}&
	\frac{1}{k}+\frac{k_x^2}{k^3} \end{pmatrix}.
\end{equation}

For steady state situations, the time partial derivative term in eq. (\ref{eq:interfvel}) vanishes and we are left with a simple self-consistent equation. In the case of a rectilinear stationary regime along the $y$-axis, $\mathbf{R}(t)=(0,Vt)^\intercal $. Eq. (\ref{eq:interfvel}) can thus be written as

\begin{equation}
\mathbf v=V\left( -\mathbf{j}+ \frac{\partial G}{\partial_y}*\boldsymbol \tau\right),
\end{equation}
where $\mathbf{j}$ is a unit vector along the $y$-axis. 

An iteration method, where the convolutions are computed in the Fourier space, is used to numerically evaluate the self-consistent solution of the problem using an empirically determined logarithmic friction law in equation~(\ref{eq:frictlaw}). The obtained velocity and stress fields are mostly homogeneous with a small distortion due to Poisson's effect. The results are very similar to the experimentally obtained fields~\cite{Nguyen2011} though no detailed comparison is presented here since the distortions are weak. 

In the case of broken lines experiments, the system is first prepared by a displacement along the $y$-axis, long enough to reach a steady state. Then, the substrate is suddenly driven along a direction which makes an angle $\theta$ with the initial displacement axis.The corresponding equation then reads

\begin{equation}
\mathbf v=-\mathbf V+\left[\frac{\partial }{\partial t}+V\left( -\sin\theta\frac{\partial}{\partial x}+\cos\theta\frac{\partial}{\partial y} \right)  \right]G*\boldsymbol \tau,
\label{eq:ligne_brisee}  
\end{equation}
where the calculated stationary state described above is used as an initial condition. This system is also solved iteratively with a time step allowing for the convergence of the solution. for the sake of simplicity, the calculations were carried out under the assumption that the contact area remains circular.  

The model presented here can be extended to more complex situations quite straightforwardly. One may describe the stress behaviour for different local friction laws: a linear dependence of the traction to the normal pressure (Coulomb's law), for example, or, for a rough glass lens on a rubber substrate, a power law~\cite{nguyen2013}. The amplitude of the interfacial stress is then determined using Hertzian stress. For smooth glass/PDMS contacts, we have also previously shown that the frictional shear stress $\tau$ is proportional to the local stretch ratio $\zeta$ within the contact, i.e. $\tau=\zeta\tau_0$, where $\tau_0$ is the stress in the absence of stretch~\cite{chateauminois2017}. Such a feature could also be implemented in the numerical resolution of the problem. For non-rubber substrate, the Poisson's ratio being different from $1/2$, normal and lateral components of stress and displacements are coupled. Though this complicates the self-consistent equations, it does not pose any particular difficulty. The occurrence of large deformations as well as non circular contact areas could also be taken into account using numerical schemes such as Finite Element (FE) simulations.\\  

It can be remarked that, for situations involving a change in the orientation of the sliding trajectory, the weak velocity dependence does not affect very much the numerical results as the amplitude of the interfacial velocity is rather homogeneous. An explicit dependence was however included in the related computations, since it does not complicates the calculation. This model is also used later in the text to account for circular trajectories or for velocity steps in linear sliding.  

%
%
\section{Discussion of the friction model for non rectilinear trajectories}
\subsubsection{Broken line trajectories}
As a validation of the friction model, we first consider the orientation of the friction force in the case of broken line sliding experiments. Numerical calculations using equation~(\ref{eq:ligne_brisee}) have been performed in situations without stick, i.e. for $\theta\leq \pi/2$. From linear sliding experiments carried out at driving velocities $V$ ranging from $1 \:\si{\micro\meter\per\second}$ to $1 \:\si{\milli\meter\per\second}$, an empirical logarithmic friction law was derived in the form
\begin{equation} 
\tau=\tau_0\left[1+k\log\left(v/v_0\right)\right],
\label{eq:friction_law}
\end{equation}
with $v_0=1 \:\si{\micro\meter\per\second}$, $\tau_0=0.163 \:\si{\mega \pascal}$ and $k=0.25$. As shown by the dotted lines in figure~\ref{fig:fig_L_force_angle}, theoretical results using this friction law are in very good agreement with experimental results. While in the point contact situation, the characteristic length involved in the reorientation of $F_T$ depends only on two geometrical parameters, namely the angle $\theta$ and the characteristic length $\lambda$, an additional length, the contact radius $a$, arises from elasticity in the case of finite size contacts. The relevance of our friction model basically depends on the ratio $\lambda/a$ which, for a pressure-independent frictional stress $\tau$, can be shown to be close to $\tau/E$. For highly rigid materials, $\tau/E \ll 1$, the reorientation of the friction force thus occurs over a very short distance. Conversely, when $\tau/E \gg 1$, the reorientation of $F_T$ obeys mainly the point contact model irrespective of the contact size. For soft materials such as elastomers, it is usually observed that $\tau/E \approx 1$~\cite{Nguyen2011,Chateauminois2008} and the orientation of the friction force then involves the sliding heterogeneities which are accounted for in the model.

\subsubsection{Circular trajectories}
\begin{figure}[!ht]
\centering
\includegraphics[width=1\textwidth]{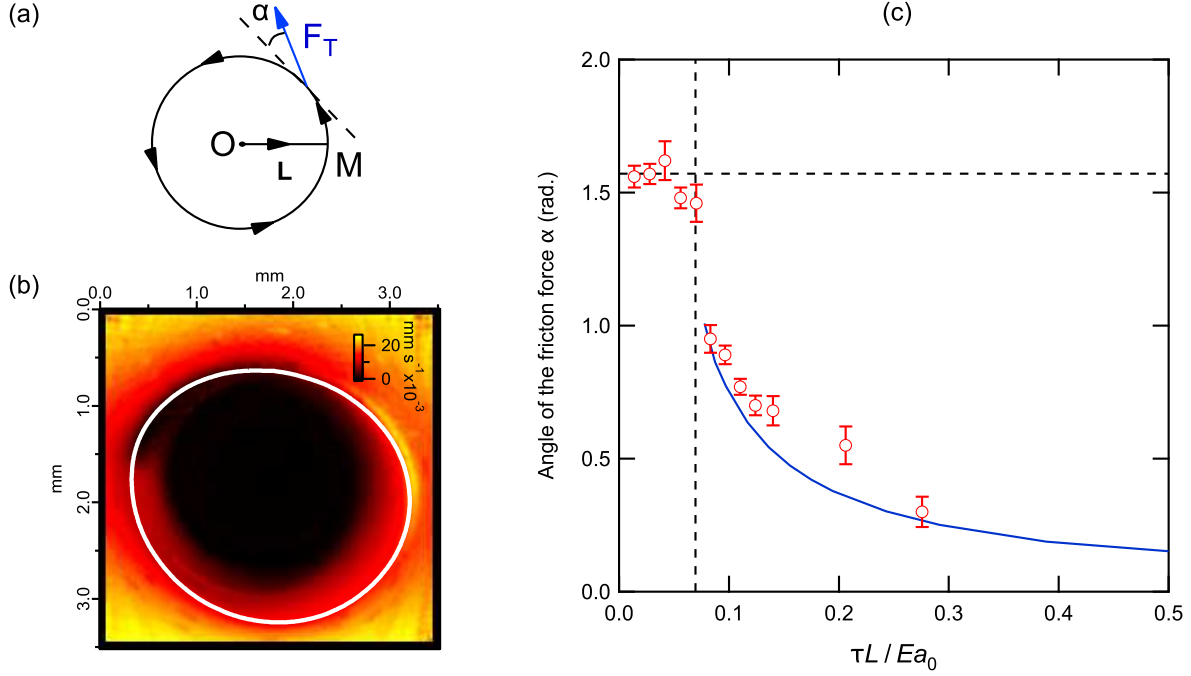}
\caption{Steady-state circular sliding trajectories. (a) Schematic of the trajectory of the PDMS substrate. Starting from rest at point O, the PDMS substrate is displaced to point $M$ where it describes a circular trajectory of radius $L$. The friction force $\boldmath{F}_T$ applied by the substrate to the lens makes an angle $\alpha$ with respect to the tangent to the circular trajectory. (b) Magnitude of the sliding velocity for $\tau L / E a_0=0.05$ ($L=0.4~\si{\milli\metre}$). The instantaneous velocity of the PDMS substrate is initially horizontal, from left to right and the centre of the circular trajectory is upwards. (c) Changes in the angle $\alpha$ as a function of the normalized radius of the trajectory $\tau L / E a_0$, where $\tau$ is the average frictional shear stress and $E$ is the Young's modulus ($V=0.04$~\si{\milli\metre\per\second}). The vertical dotted line delimits the transition from partial slip to full slip condition (for $\tau L / E a_0 \leq 0.07$) and the horizontal dotted line corresponds to $\pi/2$. The solid blue line corresponds to theoretical calculations using equation~(\ref{eq:circulaire}).}
\label{fig:circular_angle}
\end{figure}

The model was further applied to another situation where circular trajectories are applied to the PDMS substrate. Here, a steady-state situation is achieved: the substrate displacement field  is stationary in a rotating frame  where the origin of the substrate plane is fixed. By virtue of the curvature of the trajectory, there is a loss of symmetry of contact deformation. As a consequence, the friction force is no longer collinear to the tangent to the circular sliding path. As schematized in figure~\ref{fig:circular_angle}a, the PDMS substrate initially at rest is linearly displaced by a distance $L$ and then describes a circular trajectory around its initial position. Experiments have been carried out for radii of the trajectory $L$ ranging from 0.1 to 2~$\si{\milli\meter}$ (the static contact radius is $a_0=1.37 \pm 0.01 \: \si{\milli\meter}$). Here, we only consider the steady-state situations which are achieved after the substrate has described more than a whole circle. When $L < 0.5 \:\si{\milli\meter}$, a partial slip condition is evidenced from the velocity fields: as shown in the figure~\ref{fig:circular_angle}b for $L=0.4\:\si{\milli\meter}$, a large part of the contact remains stuck while some slip occurs in a crescent-like area at the contact periphery. On the other hand, a full slip condition is achieved under steady-state sliding when $L > 0.5 \:\si{\milli\meter}$.

In figure~\ref{fig:circular_angle}c, the angle $\alpha$ of the macroscopic friction force with respect to the tangent to the imposed circular motion is reported as a function of the normalized radius of the trajectory ($\tau L / E a_0$) where $\tau$ is the average frictional stress and $E$ is the Young's modulus ($\tau/E=0.19$ for the considered velocity, $V=40 \: \si{\micro\meter\per\second}$). When $L \rightarrow 0$, the size of the slip zone vanishes and the measured force then reflects the elastic response of the sheared contact. According to the loading path shown in figure~\ref{fig:circular_angle}a, the lateral force is aligned with respect to the radius of the trajectory, i.e. $\alpha=\pi/2$. Conversely, when $L \rightarrow \infty$, the curvature of the trajectories of points on the surface of the substrate becomes negligible within the contact. As a consequence, the angle of the friction force with respect to the tangent to the trajectory tends to vanish.

\begin{figure}[!ht] 
\centering
\includegraphics[width=0.5\textwidth]{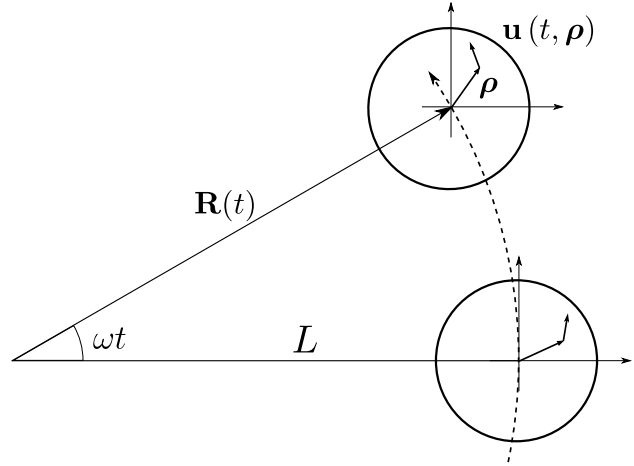}
\caption{Displacements for a circular translation of the substrate. The circles  represent the contact area at two different times and $\boldsymbol \rho $ is a given point. If rotated by $\omega t$, the substrate displacement vector, at a time $t$ for a point at $\boldsymbol\rho$ is identical to that of a point obtained by a rotation  of $-\omega t$.}%
\label{fig:rotation}%
\end{figure}

This situation was addressed theoretically by our friction model. For a steady-state circular substrate trajectory $\mathbf{R}(t)=\left( L\cos\omega t,L\sin\omega t \right)^\intercal$, the mechanical fields are stationary in a frame which is centred on the lens but which is in rotation with it (see figure \ref{fig:rotation}) i.e.
\begin{equation}
\mathbf{u}(t,\boldsymbol\rho)=\mathcal{R}_{\omega t}\mathbf{u}\left( 0,\mathcal{R}_{-\omega t}\boldsymbol\rho \right),
\end{equation} 
where $\mathcal{R}_{\alpha}$ is a rotation operator of angle $\alpha$. The time partial derivative of the displacement can then be expressed as
\begin{equation}
\frac{1}{\omega}\frac{\partial}{\partial t}\mathbf{u}(t,\boldsymbol\rho)=\mathcal{R}_{\omega t+\pi/2}\mathbf{u}\left( 0,\mathcal{R}_{-\omega t}\boldsymbol\rho \right)+\mathcal{R}_{\omega t+\pi/2}\left( \mathcal{R}_{-\omega t+\pi/2}\cdot\nabla \right)\mathbf{u}\left( 0,\mathcal{R}_{-\omega t}\boldsymbol\rho \right).
\end{equation} 
The interfacial velocity field eq. (\ref{eq:interfvel}) evaluated at $t=0$ reads
\begin{equation}
\frac{1}{\omega}\mathbf{v}=\begin{pmatrix}0\\-L\end{pmatrix}
+\begin{pmatrix}-y\partial_x+(x+L)\partial_y &-1\\1&-y\partial_x+(x+L)\partial_y\end{pmatrix}
G*\boldsymbol\tau.
\label{eq:circulaire}
\end{equation} 

In figure~\ref{fig:circular_angle}c, it turns out that the friction model captures adequately the change in the friction force angle $\alpha$ with the radius of the circular trajectory as far as a full sliding condition is considered. In its present state, our model is not able to predict the critical radius for the transition from full to partial slip. However, the impossibility of achieving convergence in the calculations close to this critical value may be viewed as a consequence of a lack of a solution ensuring a full sliding condition.

As a conclusion, the problem of circular sliding is thus driven by the ratio of the radius of the trajectory $L$ to the contact radius $a$. Any analogy with the point contact situation, \textit{i.e.} $a=0$, would therefore be inoperative. A striking feature of circular trajectories is the change in the orientation of the friction force from a radial to an orthoradial orientation when $L/a$ increases from low values in the elastic stick regime to large values in the full sliding regime.   

\subsubsection{Extension to sine wave motions}
\begin{figure}[!ht]
\centering
\includegraphics[width=0.7\textwidth]{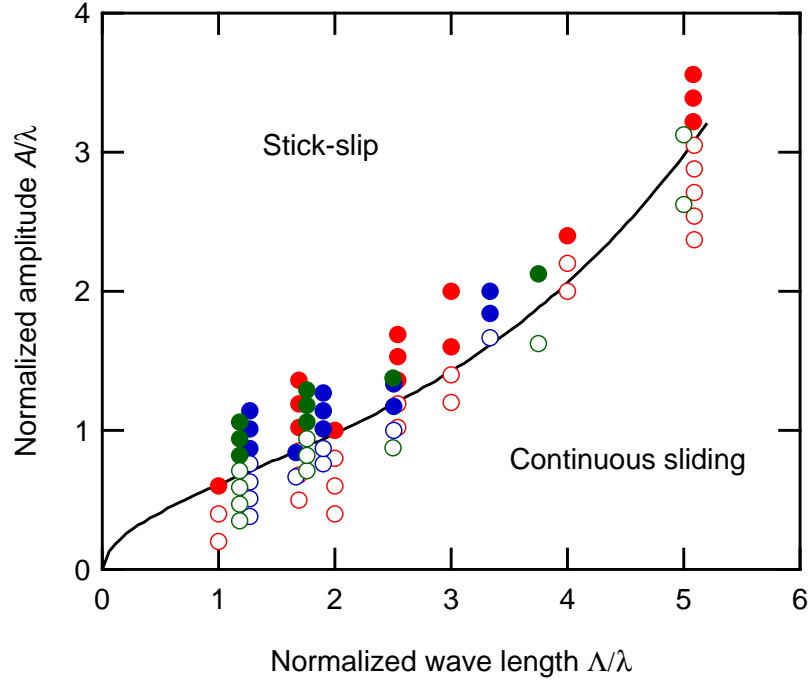}
\caption{Phase diagram for sine wave motions showing the boundary between stick-slip and continuous sliding regimes as a function of the normalized amplitude $A/\lambda$ and of the normalized wave lenth $\Lambda/\lambda$, where $\lambda$ is the elastic length  ($V=0.4 \:\si{\milli\metre\per\second}$). Experiments have been carried out with $1 \leq \Lambda \leq 6 \: \si{\milli\meter}$, $0.1 \leq A \leq 3 \:  \si{\milli\meter}$ and static contact radii $a_0$ equal to 1.3 (red), 1.7 (blue) and 2.0~$\si{\milli\meter}$ (green). Filled and open symbols correspond to data points in the stick-slip and continuous sliding regimes, respectively. The solid line is the theoretical prediction of the boundary between the two regimes using the point contact model (see Supplementary Information).}
\label{fig:phase diagram}
\end{figure}

We now extend our analysis of curvilinear sliding to sine wave motions. In such situations, the sliding interface is continuously perturbed by the reorientation of the sliding trajectory. Similarly to the broken line experiments, some stick-slip motions are induced depending on the curvature of the trajectory. In figure~\ref{fig:phase diagram}, the domain for the occurrence of such phenomena is mapped in a phase diagram as a function of the normalized amplitude $A/\lambda$ and wave length $\Lambda/\lambda$ for static contact radii ranging from 1.3 to 2.0~$\si{\milli\meter}$. From experimental data, it turns out that the boundary between the stick-slip (SS) and continuous sliding (CS) regimes is independent on the contact size. In other words, all the effects of the contact radius are embedded within the elastic length $\lambda\approx \tau a_0/E$. The independence of the boundary between CS and SS regimes on contact size is further evidenced by point contact calculations (see Supplementary Information for details): as shown by the continuous line in figure~\ref{fig:phase diagram}, these calculations provide a very good description of the CS/SS boundary for finite size contacts.

\begin{figure}[!ht]
\centering
\includegraphics[width=0.8\textwidth]{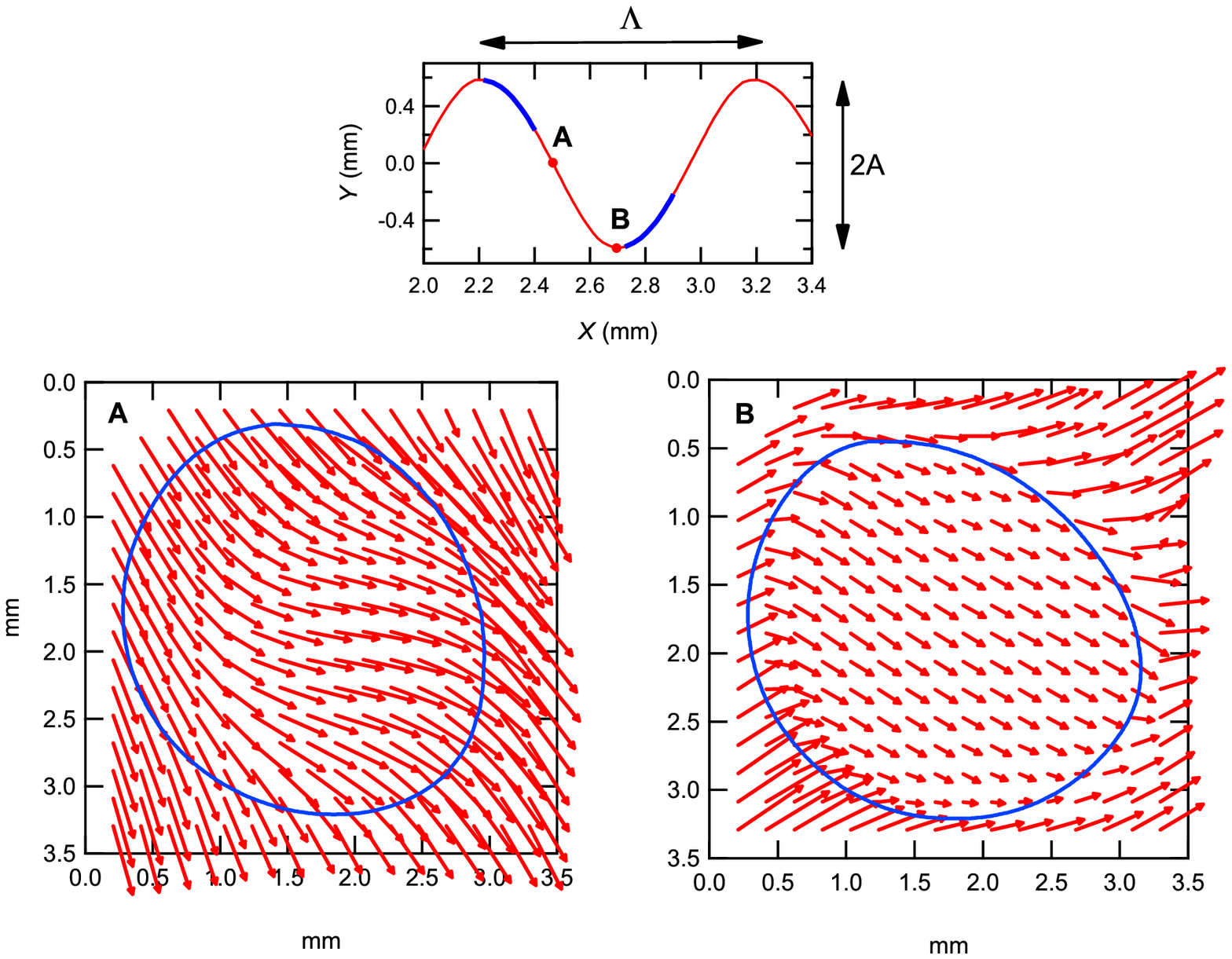}
\caption{Vector plots of the velocity field at the surface of the PDMS substrate at two positions (denoted A and B) along a sinus wave cycle in the stick-slip regime ($A=0.6\: \si{\milli\meter}$, $\Lambda=1\: \si{\milli\meter}$, $V=0.4 \:\si{\milli\metre\per\second}$). Blue lines delimit the contact area. In the top figure, bold blue lines denote the occurrence of a partial slip conditions.}
\label{fig:fields_sinus}
\end{figure}

This agreement can be rationalized from a consideration of the sliding velocity field at the transition from slip to stick conditions. As shown in figure~\ref{fig:fields_sinus}, the sliding velocity field is very homogeneous just before the contact re-stick. As a consequence, contact size is no longer a relevant length scale. Moreover, the slip to stick transition systematically occurs when the angle between the friction force and the tangent to the imposed motion is close to $\pi/2$ (results not shown), as predicted by the point contact model.

\begin{figure}[!ht]
\centering
\includegraphics[width=1\textwidth]{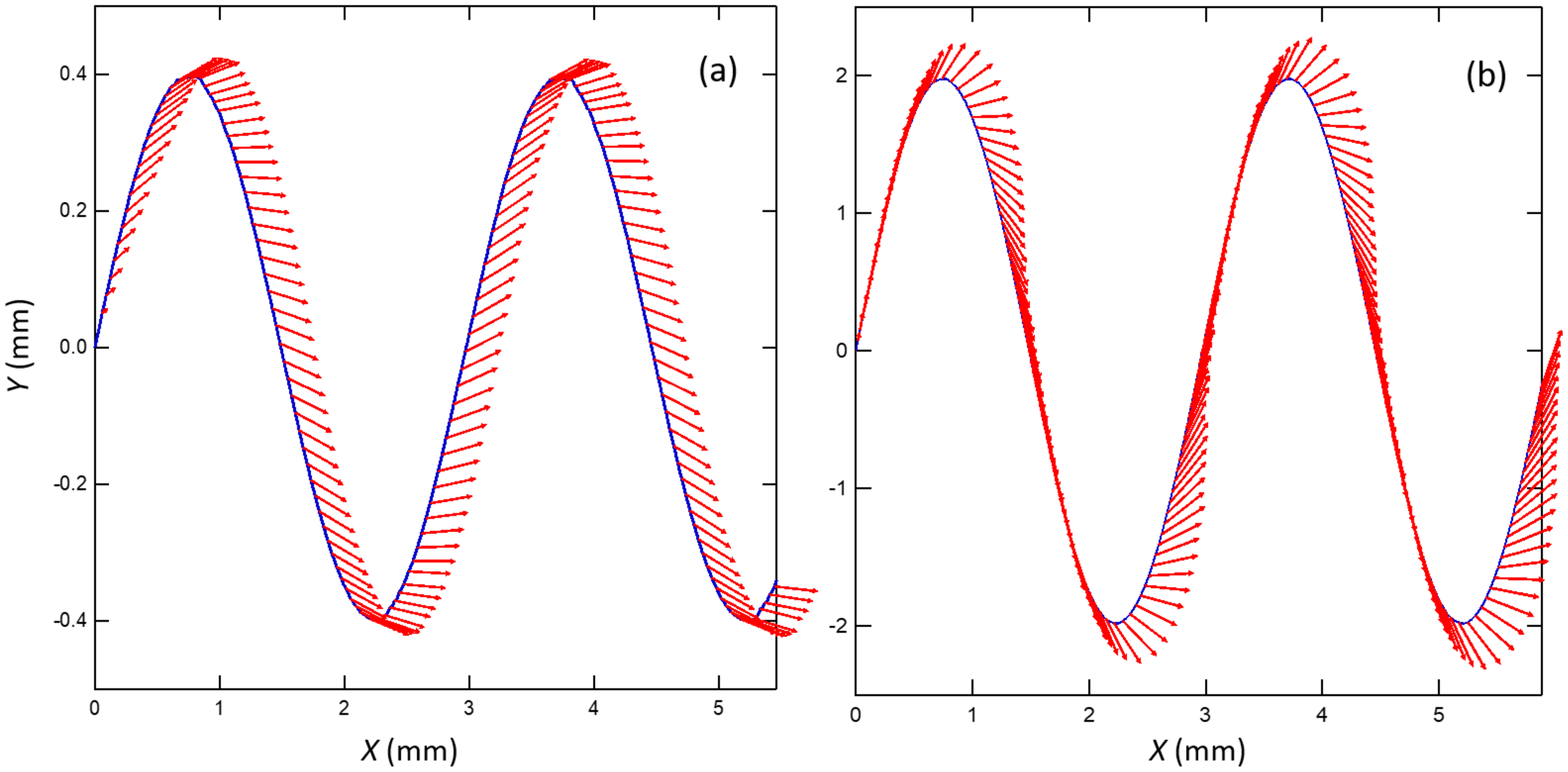}
\caption{Vector plots of the friction force applied by the PDMS substrate to the glass lens during sinus wave trajectories in the continuous sliding regime and for two values of the amplitude $A$ ($V=0.4 \:\si{\milli\metre\per\second}$, $\Lambda=3 \: \si{\milli\meter}$). (a) $A=0.4 \: \si{\milli\meter}$; (b) $A=2\: \si{\milli\meter}$.}
\label{fig:sinus_force}
\end{figure}

However, the comparison with the point contact situation fails when considering the orientation of the friction force. As shown in figure~\ref{fig:sinus_force} for experiments carried out in the continuous sliding regime, the friction force may never be aligned with respect to the tangent to the sliding trajectory of the PDMS substrate (indicated by the blue line in the figure). This is indeed the case for $A=0.4 \: \si{\milli\meter}$. For $A=2 \: \si{\milli\meter}$, the friction force only realign in a portion of the sinus cycle where the trajectory is nearly linear. As for the broken line experiments, it was observed that the orientation of the friction force corresponds to the average orientation of the sliding velocity within the contact (results not shown). Here again, the orientation of the friction force with respect to the sliding trajectory is dictated by the loss of homogeneity of contact deformation which results from the curvature of the trajectory.

%
%
\section{Velocity steps}
\begin{figure}[!h]
\centering
\includegraphics[width=0.7\textwidth]{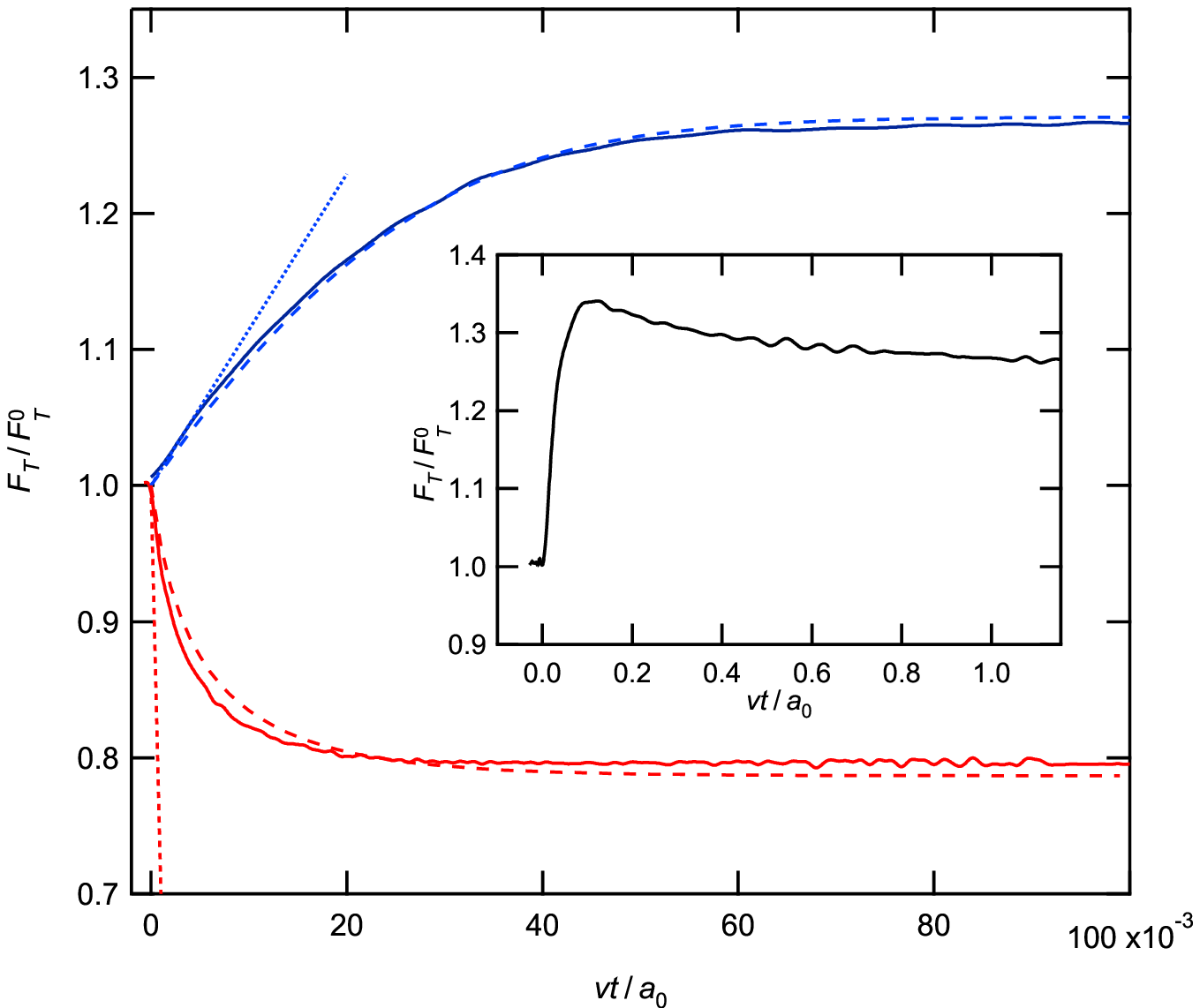}
\caption{Normalized friction force $F_T/F_T^0$ as a function of the normalized distance $Vt/a_0$ during a velocity step from 0.001 to 0.01~$\:\si{\milli\metre\per\second}$ (blue) and from 0.01 to 0.001~$\:\si{\milli\metre\per\second}$ (red). $F_T^0$ is the steady-state friction force before the application of the velocity step at $Vt/a_0=0$. Large dotted lines correspond to the simulations using equation~\ref{eq:velocity_step}. Small dotted lines correspond to equation~(\ref{eq:ponct_saut_v}) where $k=6.9\:10^3 \:\si{\newton\per\meter}$ was determined from the measured lateral contact stiffness. Inset: same for a velocity step from 0.1 to 1.0~$\:\si{\milli\metre\per\second}$, showing the occurrence of a force peak.}
\label{fig:fig_vel_step_force}
\end{figure}

We now address the generic situation of a velocity step from $V_0$ to $V$ which is often considered in the discussions of state-and-rate friction models. In figure~\ref{fig:fig_vel_step_force}, the normalized friction force $F_T/F_T^0$ (where $F_T^0$ is the steady state friction force just before the velocity step) is reported as a function of the normalized distance $Vt/a_0$. We consider here a one decade velocity step from 0.001~$\si{\milli\meter\per\second}$ to 0.01~$\si{\milli\meter\per\second}$ and, conversely, from 0.01~$\si{\milli\meter\per\second}$ to 0.001~$\si{\milli\meter\per\second}$. Here, no peak force is evidenced when the velocity is increased. The transient regime occurs over a typical distance much less than the contact size and its magnitude depends on whether the velocity is increased or decreased.

As detailed in the Supplementary Information, the transient regime can first be addressed in the light of the point contact situation. Just after the jump, calculations show that 
\begin{align}
F_T&\sim\begin{cases}
	kVt&\text{ for  }V\gg V_0\\
	-kV_0t&\text{ for }V\ll V_0  \end{cases}
\label{eq:ponct_saut_v}
\end{align}
As shown by the small dotted lines in figure~\ref{fig:fig_vel_step_force}, these point contact expressions perfectly match the initial slopes of the experimental force displacement curves just after the velocity jump.

\begin{figure}[!ht]
\centering
\includegraphics[width=0.8\textwidth]{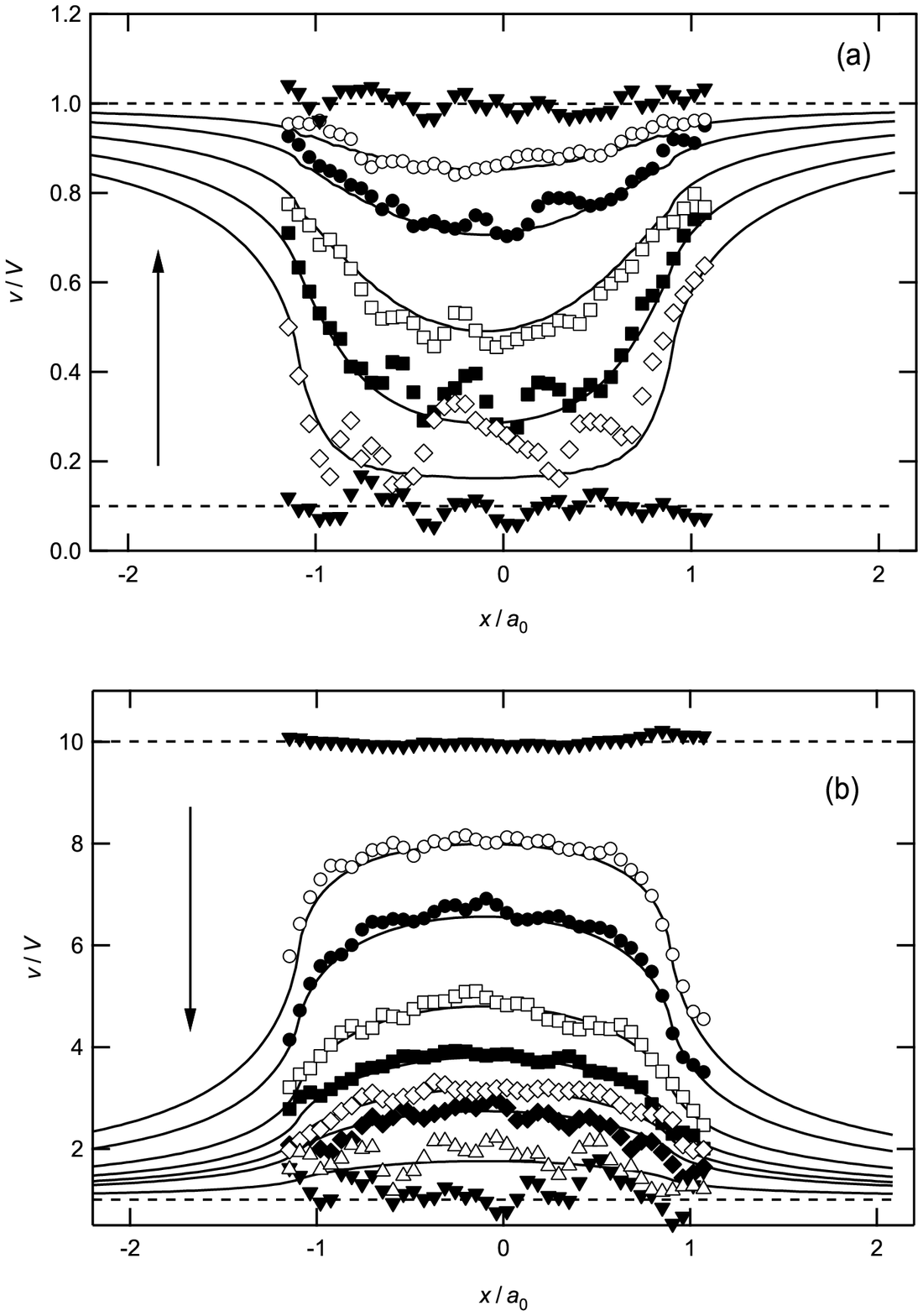}
\caption{Profiles of the normalized sliding velocity $v/V$ across the contact zone during a velocity step from (a) $V_0=0.001$ to $V=0.01\:\si{\milli\metre\per\second}$ and (b) from $V_0=0.01$ to $V=0.001\:\si{\milli\metre\per\second}$, for different normalized distances $\overline{d}=Vt/a_0$. Profiles are taken across the contact zone, normal to the direction of the sliding motions. Solid and large dotted lines correspond to the calculated profiles during the transient regime and steady-state, respectively. Open symbols correspond to experimental data.(a): $\lozenge$~$\overline{d}=0.01$, $\blacksquare$~$\overline{d}=0.02$, $\square$~$\overline{d}=0.03$, $\bullet$~$\overline{d}=0.04$, $\circ$~$\overline{d}=0.05$; (b):  $\circ$~$\overline{d}=0.0005$, $\bullet$~$\overline{d}=0.001$, $\square$~$\overline{d}=0.002$, $\blacksquare$~$\overline{d}=0.003$, $\lozenge$~$\overline{d}=0.004$, $\blacklozenge$~$\overline{d}=0.005$, $\triangle$~$\overline{d}=0.01$. $\blacktriangledown$: experimental steady-state profiles.}
\label{fig:fig_vel_step_prof}
\end{figure}

A more detailed insight into the transient regime is provided by the friction model. For a linear sliding trajectory, unsteady state situations follows from equation~\ref{eq:relative_vel}
\begin{equation}
\mathbf v=-Vj+\left( \frac{\partial}{\partial_t}+V\frac{\partial}{\partial_y}\right) G*\boldsymbol \tau,
\label{eq:velocity_step}
\end{equation}
which can be numerically solved to determine the sliding velocity field following an instantaneous change in the driving velocity from $V_0$ to $V$. In these simulations, the value $k=0.37$ of the parameter of the logarithmic friction law (equation~\ref{eq:friction_law}) was set from the experimental values of the frictional stress under steady-state sliding at $V=0.001$ and $V=0.01$~$\si{\milli\meter\per\second}$, respectively. As shown in figure~\ref{fig:fig_vel_step_force}, the simulations are in very good agreement with the experimental data. This agreement is also preserved if one consider the sliding velocity field. This is evidenced in figure~\ref{fig:fig_vel_step_prof} where experimental and calculated velocity profiles taken along contact cross-sections perpendicular to the direction of the sliding motion are reported for various non dimensional distances $Vt/a_0$. 
These profiles show that the perturbation induced by the velocity step is progressively accommodated from the periphery to the centre of the contact until steady-state is achieved, at least for the positive velocity jump.

This transient reorganization of the velocity field at the interface and the associated changes in the friction force are thus adequately described by our simple contact model which takes into account only the velocity-dependence of the frictional stress and the elastic response of the rubber substrates. Going back to the framework of the state-and-rate friction model proposed by Rice and Ruina~\cite{ruina1983,rice1983}, this means that a physical description of the state variable accounting for memory effects could tentatively be developed here from the modelling of the sliding heterogeneities induced by contact deformation.

Some limitations in this description are, however, found when the velocity step is carried out in a range of higher velocities. As shown in the inset of figure~\ref{fig:fig_vel_step_force}, when the velocity is increased from 0.1 to 1~$\si{\milli\metre\per\second}$, a peak force is induced during the transient regime which is not captured by our model. Here, some crack-like processes such as those involved in the stiction of adhesive contacts~\cite{Chateauminois2010,Barquins1975} probably occur. The description of such phenomena would deserve further investigations which are beyond the scope of this work.

\section{Conclusion}
In this study, we have addressed the issue of the transient frictional response of a glass/rubber contact interface when it is perturbed by either non rectilinear sliding motions or sliding velocity steps. The heterogeneous deformation of the interface in the finite size contact was found to control the frictional stress and the resulting macroscopic friction force. These effects were found to account for a wide variety of spectacular macroscopic behaviours which elude classical friction models. As an example, the friction force may non longer be tangent to the sliding trajectory. Some stick-slip motions induced by the curvature of the trajectories were also identified. The observed behaviour are especially relevant to soft matter systems where the ratio of the frictional stress to the Young's modulus is frequently close to unity.

In order to account for these phenomena, we have developed a simple friction model which assumes that the interfacial shear stress is oriented along the local interfacial velocity. This model was found to describe accurately the observed behaviours with non rectilinear sliding trajectories. When a velocity dependence of the frictional stress is added in the model, it also allows to account for the transient regimes resulting from velocity steps under linear sliding conditions. The observed memory effects are thus adequately explained by this model. This agreement is preserved as far as a full sliding condition is maintained at the contact interface; in its current state the model does not allow to describe partial slip situations.

The description of sliding heterogeneities at the contact scale thus provides a physical substance to the memory effects embedded in the state-and-rate friction model. This is thought to pave the way to a more physical description of several engineering problems, for which dynamic friction phenomena are involved. Bolted assemblies are for instance sometimes designed in order to ensure and localize energy dissipation in a mechanical structure such as friction dampers used as seismic resistant connections. Such bolted assemblies are also investigated for the passive control of vibrations in mechanical structures. Another class of engineering problems involving dynamic friction gathers robotic positioning applications : the control strategies usually rely on a description of the friction phenomena, which are crucial for the reliability and precision of the handling functions.

For all these applications, the fine scale description of the sliding heterogeneities is also thought to establish grounds for a macro-scale description based on yield surfaces, such as for plasticity. The simulation of complex systems involving dynamic friction could then benefit from the numerous robust numerical schemes developed for the simulation of plasticity.

\vskip6pt

\section*{Methods}
Friction experiments are carried out using transparent poly(dimethyl siloxane) (PDMS) flat substrates in contact with a smooth BK7 plano-convex glass lens (radius $R=$12.9~\si{\milli\meter}). Parallelepiped PDMS specimens 15x40x40~\si{\milli\meter\cubed} are obtained by cross-linking at 70~\si{\degreeCelsius} for 48 hours a mixture of commercially available silicone prepolymer and hardener (Sylgard 184, Dow Chemicals, USA) in a 10:1 weight ratio. As detailed in reference~\cite{Nguyen2011}, a square network of small cylindrical holes (diameter 20~\si{\micro\meter}, depth 5~\si{\micro\meter} and spacing 80~\si{\micro\meter}) is stamped on the PDMS surface by means of standard soft lithography techniques in order to measure surface displacement fields in the contact zone. Indeed, under transmitted light observation conditions, this pattern appears as a network of dark spots which are easily detected using image processing.

Friction experiments are performed using a custom-built device where a constant normal load (from 0.9 to 3.3~\si{\newton}) is applied to the glass lens by means of a dead weight arm. During experiments, the position of the glass lens is fixed while the PDMS substrate is moved by means of two crossed motorized translation stages (M.404.1PD and M.404.6PD, PI, Germany). The synchronous displacement of these two stages allows to generate various sliding trajectories (circles, broken lines or sine wave motions) at a constant imposed velocity $V$ ranging from 0.04 to 0.4~\si{\milli\metre\per\second}.
The components of the friction force $\mathbf{F}_T$ within the contact plane are continuously monitored using a dedicated, home-made,sensor located just beneath the glass lens. As fully described in~\cite{fretigny2017}, this sensor consists in a thin ($\approx 1$~\si{\milli\meter}) PDMS layer enclosed between to glass disks 20~\si{\milli\meter} in diameter. The inner faces of the two disks are patterned in order to allow for optical detection. After calibration of the isotropic shear stiffness of the sensor ($5.05 \: 10^{5}~\si{\newton\per\meter}$), the friction force components are determined from the optical measurement of the relative displacements between the two glass disks using a CCD camera (MV1-D1312, PhotonFocus, Switzerland) and a zoom lens operated in reflection mode.
Pictures of the contact are continuously recorded through the thickness of the transparent PDMS substrate using a CCD camera  (2048 x 2048$^{2}$, 8 bits, SVS Exo,Vistek, Germany) and a long-working distance objective (APO Z16, Leica, France). Displacement fields are measured with sub-pixel accuracy from the detection of the dots pattern on the PDMS substrate using conventional image processing techniques. They were systematically corrected from the measured displacements of the lens which result from the compliances of the dead weight arm supporting the lens and of the load sensor.

Velocity steps experiments were carried out using a separate, dedicated, linear sliding device equipped with a high bandwidth (5~$\si{kHz}$) lateral load sensor (Sensotec 5N, Spain) and a data acquisition system operated up to a 1~$\si{kHz}$ sampling rate. Experiments were carried out at imposed velocity under a constant 1~\si{\newton} normal force applied using a linear coil actuator (PIMag V-275, PI, Germany) in closed loop control. The sliding velocity was varied from 0.001 to 1.0~\si{\milli\metre\per\second} by means of a motorized linear translation stage (M403.4PD, PI, Germany). The ratio of the lateral device stiffness to the lateral contact stiffness is greater than 50. Contact pictures were recorded through the PDMS substrate using the same optics as that described above and a camera ($1024^2$,8 bits, MV-D1024, PhotonFocus, Switzerland) which was operated up to a frame rate of 90~$\si{kHz}$.\\
\bibliographystyle{RS} 

\newpage

\begin{figure}[h]
	\centering
	\includegraphics[width=0.7\textwidth]{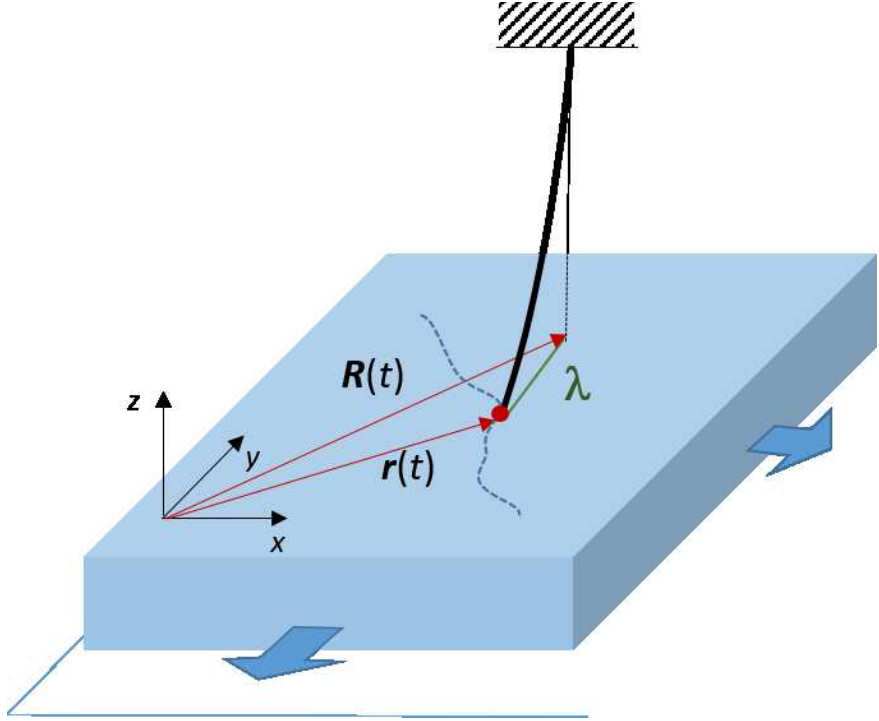}
	\caption{Schematic representation for a possible realisation of a point contact (red bullet) lying on a moving flat substrate. The point contact is fixed to the frame of the laboratory by means of a flexible fibre with a two-dimensional isotropic compliance in the substrate plane. $\mathbf{r}(t)$ and $\mathbf{R}(t)$ denote the locations of the point contact and of the projection of the holder, respectively, in a frame $(x,y)$ attached to the moving substrate.  In full sliding conditions, the extension of the fibre corresponds to the tribo-elastic length $\lambda$ and the direction of the projection of the fibre is tangent to the slider trajectory (indicated by dotted lines).}
	\label{fig:schematic_contponct}%
\end{figure}

\begin{center}
	\bigskip{Supplementary Information, Fazio \textit{et al}}\\
\end{center}
\begin{center}
	\section*{	
		\Large {Point contact in the sliding regime: generalised tractrix}}
\end{center}
\renewcommand{\theequation}{SI. \arabic{equation}}
As schematically depicted in figure~\ref{fig:schematic_contponct}, a point contact (red bullet) is lying on a moving substrate and held to a fixed holder by means of a flexible fibre with an 2D isotropic compliance $k$ in the plane of the substrate. In the $(x,y)$ plane, the location of the point contact and of the vertical projection of the holder are denoted $\mathbf{r}(t)=(x(t),y(t))^\intercal$ and $\mathbf{R}(t)=(X(t),Y(t))^\intercal$, respectively. The distance in between these two points is the tribo-elastic length $\lambda$ and the line passing through both points is tangent to the slider trajectory. These conditions define this trajectory as a generalised tractrix corresponding to the directrix $\mathbf R(t)$ and to the parameter $\lambda$ (see for example reference~\cite{cady1965})\footnote{From reference \cite{cady1965}: ``\textit{At a meeting in Paris in 1693 Claude Perrault laid his watch on the table, with the long chain drawn out in a straight line... He showed that when he moved the end of the chain along a straight line, keeping the chain taut, the watch was dragged along a certain curve. This was one of the early demonstrations of the tractrix. The line along which the chain is pulled is called the directrix.}'' In the present work, the situation is analogous, the role of the chain being played by the tribo-elastic length $\lambda$ }. It obeys the following equations:
\begin{align}
	x(t)+\frac{\lambda \dot x(t)}{\sqrt{\dot x^2(t)+ \dot y^2(t)}}&=X(t)\nonumber\\
	y(t)+\frac{\lambda \dot y(t)}{\sqrt{\dot x^2(t)+\dot y^2(t)}}&=Y(t)\label{eq:tractGen}
\end{align}
An initial condition must be given, where the slider-holder distance is $\lambda$. As discussed in the main text, this sliding regime exists if $\dot{\mathbf{R}}(t)(\mathbf{R}(t)-\mathbf{r}(t))=\dot{\mathbf{R}}(t)\dot{\mathbf{r}}(t)>0$. 
When this condition fails at $t=t_s$, the slider stops with respect to the substrate until the sliding condition $\left|r(t_s)-R(t)\right|=\lambda$ becomes fulfilled again. The trajectory is then a tractrix again with a new  initial condition.\\

\subsection*{Linear displacement}
For a linear displacement of the substrate along the $x$-axis, then, $X(t)=V t,Y(t)=0$, it can be checked that the general solution of this system verifying $x(0)=\lambda\cos\varphi,\,y(0)=\lambda\sin\varphi$ with $\pi/2\le\varphi\le\pi$ can be written as 
\begin{align}
	x(t)&=Vt-\lambda\left[\tanh\left( \frac{V(t+t_0)}{\lambda} \right)-\tanh\left( \frac{Vt}{\lambda} \right)\right]-\lambda\tanh\left( \frac{Vt_0}{\lambda} \right)\\
	y(t)&=\frac{\lambda}{\cosh\left( \frac{V(t+t_0)}{\lambda} \right)}
\end{align}
with
\begin{align}
	\cos\varphi=-\tanh\left( \frac{Vt_0}{\lambda} \right);\hspace{1cm}\sin\varphi=\frac{1}{\cosh\left( \frac{Vt_0}{\lambda} \right)}
\end{align}
With adimensionalised coordinates $\bar x(t)=x(t)/\lambda,\bar y(t)=y(t)/\lambda$, with $u=Vt/\lambda$, the parametric equations read
\begin{align}
	\bar x(u)&=u-\frac{\tanh u-\cos\varphi}{1-\cos\varphi\tanh u}\\
	\bar y(u)&=\frac{\sin\varphi}{\cosh u-\sinh u\cos\varphi}
\end{align} 

To describe an experiment where the driving direction is suddenly reoriented by an angle $\theta$ at $t=0$ after a sliding phase along the $x$-axis, the previous trajectory must be rotated by an angle $\theta$ around the origin. Two situations may occur:
\begin{itemize}
	\item When $\theta<\pi/2$, there is not stick-phase. For $t>0$ the trajectory of the slider follows a tractrix curve passing through $F$. The initial condition at $u=0$ is $\varphi=\pi-\theta$ (see Fig. 1a in the main text).
	\item When $\theta>\pi/2$, the slider stops at $t=0$ while the driver still moves until the distance between both points reaches $\lambda$ again, at an instant $t_0=-2(\lambda/V)\cos(\theta)$. This condition corresponds to $\varphi=\theta$ (see Fig. 1b in the main text). 
\end{itemize}
\subsection*{Sine driving curve}
For a general driving curve, depending on the geometrical characteristics of the curve, the slider may follow pieces of generalised tractrices during the slip phases or remain motionless on the substrate during the stick phases. 

For a sine driving curve ($X(t)=vt,Y(t)=A \sin(2 \pi vt/\Lambda)$), the steady state regime may correspond to continuous sliding or periodic stick-slip. The  trajectory is numerically determined, by solving eq. (\ref{eq:tractGen}) in the sliding phases. For a given period, by calculating the solution for increasing values of the amplitude, the boundary of the steady sliding regime is found when in the established regime the quantity $d\mathbf{R}/dt\left(\mathbf{R}(t)-\mathbf{r}(t)\right)$ becomes negative at some point. If this quantity remains positive, no stick phases occurs and the numerically obtained solution is valid. If it changes its sign at $t=t_s$, the slider stops ($\dot{\mathbf{r}}(t)=0$) until $\left|\mathbf{r}(t_s)-\mathbf{R}(t)\right|$ reaches the value $\lambda$ again, giving a new initial condition for the eq. (\ref{eq:tractGen}). The boundary domain obtained in this way is reported in figure 8 in the main text for comparison to the finite contact size case.

\subsection*{Velocity step}
We consider here the response of the point contact to a sudden jump in the sliding velocity, from $V$ to $V_0$ at time $t=0$. The following velocity-dependent friction law is assumed
\begin{align}
	F_T=A+B\ln \frac {V}{V_0}
\end{align}
where $A$ and $B$ are two constants and $V_0$ is the initial velocity. This velocity step from $V_0$ à $V$ thus corresponds to a change in the elastic length from $\lambda_0=A/k$ to $\lambda=\lambda_0+B/k\ln V/V_0$. The motion obeys the following equation
\begin{align}
	-k(x(t)-vt)=A+B\ln\left(\frac{\dot x(t)}{v_0}\right)
\end{align}
The solution $x(t)=-\frac Ak+ \frac Bk\ln\left[1+\frac{V_0}V\left(e^{\frac{kVt}B}-1\right)\right]$ gives the friction force
\begin{align}
	F_T=A+B\ln\left( \frac{V}{V_0} \frac{1}{1+e^{-\frac {Vt}\delta}\left( \frac{V}{V_0}-1 \right)}\right)
\end{align}
with $\delta =B/k$. 
\begin{figure}[h]
	\centering
	\includegraphics[width=0.7\textwidth]{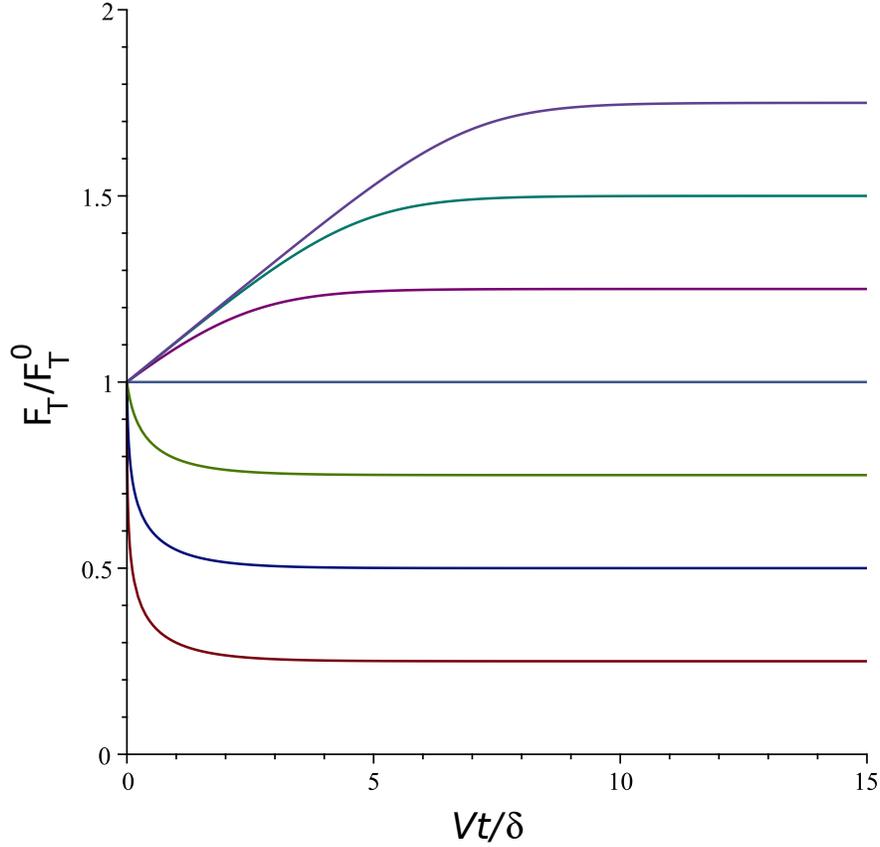}
	\caption{Change in the normalized friction force $F_T/F_T^0$ as a function of the normalized displacement of the driving system $Vt/\delta$ for velocity steps with magnitudes from $10^{-3}$ to $10^3$ by one decade increments. ($A=0.163,B=0.041,V_0=10^{-6}$).}%
	\label{fig:steps}%
\end{figure}
The calculated friction force is reported as a function of the normalized driving displacement $Vt/\delta$ in figure \ref{fig:steps} for velocity steps from $10^{-3}$ to ${10^3}$ by one decade increments. For each increase in velocity, an unique slope is achieved at short times whatever the magnitude of the velocity step. The same is observed for a decrease in velocity but with a steeper slope. Accordingly, the calculation shows that just after the jump
\begin{align}
	T&\sim\begin{cases}
		kVt&\text{ pour }V\gg V_0\\
		-kV_0t&\text{ pour }V\ll V_0  \end{cases}
\end{align}

\end{document}